\documentclass{JHEP3}
\usepackage{graphicx}
\def\be{\begin{eqnarray}}\def\ba{\begin{eqnarray}}
\def\ee{\end{eqnarray}}\def\ea{\end{eqnarray}}

\title{Quark-gluon mixed condensate of the QCD vacuum in Holographic QCD}

\author{Hyun-Chul Kim \\ Department of Physics,
Inha University, Incheon 402-751, Korea \\
E-mail: \email{hchkim@inha.ac.kr} }
\author{Youngman Kim \\ School of Physics, Korea Institute for Advanced
Study, Seoul 130-012, Korea \\
E-mail: \email{chunboo81@kias.re.kr}}
\abstract{We investigate the quark-gluon mixed condensate based on
an AdS/QCD model.  Introducing a holographic field dual  to the
operator for the quark-gluon mixed condensate, we obtain the
corresponding classical equation of motion. Taking the mixed condensate as an additional
free parameter, we show that the present scheme reproduces very well
experimental data. A fixed value of the mixed condensate is in good
agreement with that of the QCD sum rules.
}

\preprint{INHA-NTG-13/2008}

\begin{document}

\section{Introduction}
The QCD vacuum is known to be very complicated due to both
perturbative and non-perturbative fluctuations.  In particular, the
quark and gluon condensates, being the lowest dimensional ones,
feature the non-perturbative structure of the QCD vacuum.  The quark
condensate can be identified as the order parameter for the
spontaneous breakdown of chiral symmetry (SB$\chi$S) which
plays an essential role in describing low-energy phenomena of hadrons:
In the QCD sum rule, the chiral condensate arises from the operator
product expansion and is determined phenomenologically by hadronic
observables~\cite{Shifman:1978bx,Shifman:1978by}, while in chiral
perturbation theory ($\chi$PT), it is introduced in the mass term of
the effective chiral Lagrangian at the leading
order~\cite{Gasser:1983yg,GellMann:1968rz}.

While the quark and gluon condensates are well understood
phenomenologically, higher dimensional condensates suffer from
large uncertainty.  Though it is possible to estimate
dimension-six four-quark condensates in terms of the quark condensate
by using the factorization scheme that is justified in the large
$N_c$ limit, the dimension-five mixed quark-gluon condensate is not
easily determined
phenomenologically~\cite{Ioffe:1981kw,Belyaev:1982sa,
Dosch:1988vv,Dorokhov:1997iv, Polyakov:1996kh,Doi:2002wk,Nam:2006ng}.
In particular, the mixed condensate is an essential parameter to
calculate baryon masses~\cite{Ioffe:1981kw}, exotic hybrid
mesons~\cite{Balitsky:1986hf}, higher-twist meson distribution
amplitudes~\cite{Chernyak:1983ej} within the QCD sum rules.
Moreover, the mixed quark condensate can be regarded as
an additional order parameter for the SB$\chi$S since the quark
chirality flips via the quark-gluon operator.  Thus, it is naturally
expressed in terms of the quark condensate:
\begin{equation}
\langle\bar{\psi}\sigma^{\mu\nu}G_{\mu\nu }
\psi\rangle = m_0^2 \langle\bar{\psi}\psi\rangle
\label{eq:mc1}
\end{equation}
with the dimensional parameter $m_0^2$ which was estimated in various
works~\cite{Ioffe:1981kw,Belyaev:1982sa,Dosch:1988vv,Dorokhov:1997iv,
Polyakov:1996kh,Doi:2002wk,Nam:2006ng}. The gluon-field strength is defined as
$G_{\mu\nu}=G^a_{\mu\nu}\lambda^a/2$.

The AdS/CFT
correspondence~\cite{Maldacena:1997re,Gubser:1998bc,Witten:1998qj}
that provides a connection between a strongly coupled large $N_c$
gauge theory and a weakly coupled supergravity gives new and
attractive insight into nonperturbative features of quantum
chromodynamics (QCD) such as the quark confinement and spontaneous
breakdown of chiral symmetry (SB$\chi$S).  Though there is still no
rigorous theoretical ground for such a correspondence in real QCD,
this new idea has triggered a great amount of theoretical works on
possible mappings from nonperturbative QCD to $5D$ gravity,
i.e. holographic dual of QCD.  In fact, there are in general two
different routes to modeling holographic dual of QCD (See, for
example, a recent review~\cite{Erdmenger:2007cm}): One way is to
construct 10 dimensional (10D) models based on string theory of D3/D7,
D4/D6 or D4/D8 branes~\cite{Karch:2002sh,Kruczenski:2003be,
Kruczenski:2003uq,Sakai:2004cn,Sakai:2005yt}.  The other way is
so-called a bottom-up approach to a holographic model of
QCD, often called as
AdS (Anti-de Sitter
Space)/QCD~\cite{Erlich:2005qh,DaRold:2005zs,DaRold:2005vr} in which a
5D holographic dual is constructed from QCD, the 5D gauge coupling
being identified by matching the two-point vector correlation
functions.  Despite the fact that this bottom-up approach is somewhat
on an ad hoc basis, it reflects some of most important features of
gauge/gravity dual. Moreover, it is rather successful in describing
properties of hadrons (See the recent
review~\cite{Erdmenger:2007cm}).

In the present work, we aim at investigating the mixed condensate of
the QCD vacuum defined as Eq.(\ref{eq:mc1}), closely following
ref.~\cite{Erlich:2005qh}.  The hard-wall model of
ref.~\cite{Erlich:2005qh} is the simplest one but provides a clean-cut
framework to study the mixed condensate.  Thus, we will extend the 5D
action in ref.~\cite{Erlich:2005qh}, introducing the bulk field
corresponding to the operator for the mixed condensate. We will
carefully examine how the meson masses and couplings undergo changes
in the presence of the mixed condensate.

We sketch the present work as follows: In section 2, we describe
briefly the hard-wall AdS/QCD model with the bulk field for the mixed
condensate taken into account.  In section 3, we show our
results for the mixed condensate. we also examine, in the presence of
the mixed condensate, possible changes of the meson observables such as
masses and coupling constants. In the last section, we summarize the
present work and draw conclusions.

\section{A hard-wall AdS/QCD model}
The metric of an AdS space is given as
\begin{equation}
ds^2\;=\;g_{MN}dx^M dx^N \;=\;\frac1{z^2}(\eta_{\mu\nu} dx^\mu dx^\nu -
dz^2),
\end{equation}
where $\eta_{\mu\nu}$ denotes the 4D Minkowski metric:
$\eta_{\mu\nu}=\mathrm{diag}(1,\,-1,\,-1,\,-1)$. The AdS space is
compactified by two different boundary conditions, i.e. the IR
bounadry at $z=z_m$, and the UV at $z=\epsilon\to 0$. Thus, the model
is defined within the range: $\epsilon\le z \le z_m$.
Taking into account the bulk field corresponding to the operator for
the mixed condensate, we express the classical 5D bulk action as
follows:
\begin{equation}
S\;=\;\int\,d^5x\,\sqrt{g}\,\mathrm{Tr}\left[|DX|^2+3|X|^2 -
  \frac1{4g_5^2}(F_L^2+F_R^2) + |D\Phi|^2 - 5\Phi^2
\right] ,
\end{equation}
where $D_\mu X\;=\;\partial_\mu X-iA_{L\mu}X + iXA_{R\mu}$ and
$F_{L,R}^{\mu\nu}\;=\; \partial^\mu A_{L,R}^\nu-\partial^\nu
A_{L,R}^\mu -i[A_{L,R}^\mu,\,A_{L,R}^\nu]$. The massless gauge field
is defined as $A_{L,R} =A_{L,R}^{a}t^a$ with
$\mathrm{tr}(t^at^b)=\delta^{ab}/2$. The 5D masses $m_5^2$ of the bulk
fields are determined by the relation
$m_5^2\;=\;(\Delta-p)(\Delta+p-4)$~\cite{Gubser:1998bc,Witten:1998qj},
where $\Delta$ stands for the dimension of the corresponding operator
with spin $p$.  In table~\ref{tab:1}, the operators and corresponding
bulk fields are listed with the 5D masses given.
\begin{table}[ht]
  \centering
  \begin{tabular}{lcccc}
\hline\hline
4D operators: $\mathcal{O}(x)$ & 5D fields: $\phi(x,z)$ & $p$ &
$\Delta$ & $m_5^2$  \\ \hline
$\bar{q}_L\gamma^\mu t^a q_L$ & $A_{L\mu}^a$ & 1 & 3 & 0 \\
$\bar{q}_R\gamma^\mu t^a q_R$ & $A_{R\mu}^a$ & 1 & 3 & 0 \\
$\bar{q}_R^\alpha q_L^\beta$ & $(2/z)X^{\alpha\beta}$ & 0 & 3 & -3 \\
$\bar{q}_R^\alpha\sigma_{\mu\nu}G^{\mu\nu} q_L^\beta$ &
$(1/z^3)\Phi^{\alpha\beta}$ & 0 & 5 & 5 \\
\hline\hline
  \end{tabular}
  \caption{A dictionary for 4D operators and 5D fields}
  \label{tab:1}
\end{table}
$g_5$ represents the 5D gauge coupling.  The bi-fundamental scalar
field $X$ is relevant to SB$\chi$S.  Its vacuum expectation value
(VEV) is holographic dual to the bilinear scalar quark operator
$\bar{q}_Lq_R$, which can be written in terms of the chiral condensate
$\langle \bar{q}_Lq_R\rangle = \Sigma =\sigma\mathbf{1}$ and the
current quark mass
$\hat{m}=\mathrm{diag}(m_{\mathrm{q}},\,m_{\mathrm{q}})$.  The
bi-fundamental scalar field $\Phi(x,z)$ is introduced as a holographic
dual for the operator of the mixed quark-gluon condensate $\bar{q}_R
\sigma_{\mu\nu}G^{\mu\nu} q_L$. In this work, we focus on the VEV of $\Phi (x,z)$
to study the mixed condensate.

In ref.~\cite{Klebanov:1999tb}, it was shown
that for small $z$ or near the boundary of AdS space, a 5D field
$\phi(x,z)$ dual to a 4D operator $\mathcal{O}$ can be expressed as
\begin{equation}
\phi(x,z)\;=\; z^{4-\Delta}  \left[\phi_0(x) + O(z^2)\right]+z^\Delta
\left[A(x) + O(z^2)\right],
\end{equation}
where $\phi_0(x)$ is a prescribed source function for $\mathcal{O}(x)$
and $A(x)$ denotes a physical fluctuation that can be
determined from the source by solving the classical equation of
motion.  It can be directly related to the VEV of the $\mathcal{O}(x)$
as follows~\cite{Klebanov:1999tb}:
\begin{equation}
A(x)\;=\;\frac1{2\Delta -4} \langle\mathcal{O}(x)\rangle.
\end{equation}
Thus, from the classical equations of motion of
$X$~\cite{Erlich:2005qh, DaRold:2005zs} and $\Phi$, we obtain
\begin{eqnarray}
X_0 (x,z) &=& \langle X(x,z)\rangle\;=\;  \frac12 (\hat{m} z + \sigma
z^3) ,\cr
\Phi_0(x,z) &=& \langle \Phi(x,z)\rangle\;=\; \frac16(c_1 z^{-1}
+\sigma_M z^5 ),
\end{eqnarray}
where $c_1$ is the source term for the mixed condensate and
$\sigma_M$ represents the mixed condensate
$\sigma_M\;=\;\langle\bar{q}_R \sigma_{\mu\nu}G^{\mu\nu} q_L\rangle$.
For simplicity, we take $c_1=0$.

We now review how to fix the 5D gauge
coupling~\cite{Erlich:2005qh,DaRold:2005zs} by matching the
two-point vector correlation function to the leading contribution from
the OPE result~\cite{Shifman:1978bx}.  The vector field $V$ is defined
as $(A_L+A_R)/2$ with the axial-like gauge condition
$V_z(x,z)\;=\;0$.  It can be decomposed into the transverse and
longitudinal parts: $V_\mu\;=\; (V_\mu)_\perp +(V_\mu)_\|$.
Using the Fourier transform of the vector field: $V_\mu^a=\int d^4x
e^{iq\cdot x} V_\mu^a(x,z)$, we can write the equation of motion for
the transverse part of the vector field as
follows~\cite{Erlich:2005qh}:
\begin{equation}
\left[\partial_z\left(\frac1{z}\partial_z V_\mu^a(q,\,z)\right) +
  \frac{q^2}{z} V_\mu^a(q,\,z) \right]_\perp \;=\;  0.
\label{eq:transv}
\end{equation}
The corresponding solution of Eq.(\ref{eq:transv}) can be expressed as
a separable form:
\begin{equation}
(V_\mu^a(q,\,z))_\perp \;=\; V(q,\,z) \overline{V}_\mu^a(q),
\end{equation}
where $\overline{V}_\mu^a (q)$ is the Fourier transform of the source
of the 4D conserved vector current operator $\bar{q}\gamma_\mu t^a q$
at the boundary.  The explicit solution for $V(q,\,z)$ can be derived
by solving Eq.(\ref{eq:transv}) with the boundary conditions
$V(q,\,\epsilon)\;=\;1$ and $\partial_zV(q,\,z_m)\;=\;0$:
\begin{equation}
V(q,z)\;=\;\frac{\pi q z}{2}\left[\frac{Y_0(q z_m)}{2 J_0(q
    {z_m})}J_1(q z)
- Y_1(q z)\right],
\end{equation}
where $J_i$ and $Y_i$ denote the Bessel functions of the first and
second kinds, respectively.  The $V(q,\,z)$ is often called a
bulk-to-boundary propagator, since the solution $V(q,z)$ leaves only
the boundary term of the action in a quadratic form:
\begin{equation}
S_{\mathrm{b}}\;=\; -\frac1{2g_5^2}\int d^4
x\overline{V}_\mu(q)\left(\frac{\partial_z V(q,z)}{z} \right)_{z=\epsilon}
\overline{V}^\mu(q) .
\end{equation}
Thus, the correlation function can be obtained by the second
derivative of the action with respect to the vector field
$\overline{V}^\mu(q)$:
\begin{equation}
\Pi_V(Q^2) \;=\; -\frac{1}{g_5^2 Q^2}\frac{\partial_z V(q,\,z)}{z},
\end{equation}
where $Q^2=-q^2$.  In a large Euclidean region ($Q^2\to\infty$), one
gets
\begin{equation}
\Pi_V(Q^2) \;=\; -\frac1{2g_5^2} \ln Q^2.
\end{equation}
Since it is already known from the OPE that the vector correlation
function in the leading order is given as
\begin{equation}
\Pi_V(Q^2) \;=\; -\frac{N_c}{24\pi^2} \ln Q^2,
\end{equation}
one can immediately determine the 5D gauge coupling $g_5^2$:
\begin{equation}
g_5^2\;=\; \frac{12\pi^2}{N_c}.
\end{equation}

We are now in a position to derive the classical equations of motion
for the axial-vector and pion.
Introducing $v=m_qz +\sigma z^3$,
$w = (\sigma_M/3) z^5 $, and $(A_\mu)_\| =\partial_\mu\phi$, we obtain
\begin{eqnarray}
  \label{eq:7}
&& \left[\partial_z \left(\frac1{z}\partial_z A_\mu\right) +\frac{q^2}{z}
   A_\mu   - g_5^2\frac{1}{z^3} (v^2+w^2) A_\mu\right]_\perp \;=\; 0\,
 , \label{EqA}\\
 && \partial_z\left ( \frac{1}{z}\partial_z\phi^a\right) +
 g_5^2\frac{1}{z^3}v^2 (\pi^a-\phi^a)=0\, ,\cr
 && -q^2\partial_z\phi^a +g_5^2\frac{1}{z^2}
 (v^2+w^2)\partial_z\pi^a=0\, .
\end{eqnarray}
Finally, we consider decay constants and  interactions in the
model~\cite{Erlich:2005qh,DaRold:2005zs}.
The decay constant of $\rho$ is given by
\begin{equation}
F_\rho^2=\frac{1}{g_5^2}\left(\frac{\psi^\prime_\rho
    (\epsilon)}{\epsilon} \right)^2\, ,
\end{equation}
where $\psi_\rho (z)$ denotes a $\rho$-meson wave function defined as:
$V_\mu(x,z)=g_5\sum_n V_\mu^{(n)}(x)\psi^{(n)}(z)$.  The  $\rho$-meson
wave function is the solution of (\ref{eq:transv}) at
$q^2=m_\rho^2$ with the boundary conditions $\psi_\rho (\epsilon)=0$
and $\partial_z\psi_\rho (z_m)=0$ imposed.  The pion decay constant is
\begin{equation}
f_\pi^2=\left. -\frac{1}{g_5^2} \frac{\partial_z
    A(0,z)}{z}\right|_{z=\epsilon}\, ,
\end{equation}
where $A(0,z)$ is the solution of (\ref{EqA}) with $q^2=0$ and with
two boundary conditions:
$A(0,\epsilon)=1$, $A^\prime (0,z_m)=0$.
The $\pi-\rho$ coupling reads as
follows~\cite{Erlich:2005qh,DaRold:2005zs}:
\begin{equation}
g_{\rho\pi\pi} =g_5\int_\epsilon^{z_m} dz \psi_\rho(z)\left(
  \frac{{\phi^\prime}^2}{g_5^2z} +
  \frac{v(z)^2(\pi-\phi)^2}{z^3}\right) \, .
\end{equation}
\section{Numerical Results}
In this section,  we present the numerical results of various hadronic
observables and condensates discussed previously.

%%%
\begin{table}[ht]
  \centering
\begin{tabular}{cccccc}
\hline\hline
   &{\rm Model I } &{\rm Model II} &{\rm Model III }&{\rm Model
     IV}&{\rm Experiment} \\
  \hline
 $m_{\mathrm{q}}$ & $1.6$&$3.7$ &$2.3$ &$2.3$ & $\cdots$ \\
  $\sigma$ & $(0.1\,\mathrm{GeV})^3$ & $(0.25\,\mathrm{GeV})^3$ &
  $(0.307\,\mathrm{GeV})^3$ & $(0.308\,\mathrm{GeV})^3$ &$\cdots$ \\
$m_0^2$ & $13.32\,\mathrm{GeV}^2$ & $0.72\,\mathrm{GeV}^2$ &
$0.006\,\mathrm{GeV}^2$ & $0$ & $\cdots$\\
$m_\rho$ & $775.8$ & $775.8$ & $775.8$ &$832$ & $775.49\pm 0.34$  \\
$m_{a_1}$ & $1230$ & $1244$ & $1246$ &$1220$ & $1230\pm 40$ \\
$f_\pi$ & $75.9$ & $80.5$ & $85.5$ & $84.0$ & $92.4\pm 0.35$  \\
$F_\rho^{1/2}$ & $330$ & $330$ & $330$ & $353$ & $345\pm 8$  \\
$F_{a_1}^{1/2}$ & $460$ & $459$ & $446$ & $440$ & $433\pm 13$ \\
$m_\pi$ & $138$ & $139.3$ & $137.5$ & $141$ & $139.57\pm 0.00035$ \\
$g_{\rho\pi\pi}$ & $8.27$ & $4.87$ & $4.87$ & $5.29$ & $6.03\pm 0.07$ \\
$g_{A4}$& $1.71$ & $1.69$ & $1.71$ & $1.88$ &$\cdots$ \\ \hline \hline
\end{tabular}
  \caption{The results of the model with and without the mixed
condensate. Model IV corresponds to Model B of
ref.~\cite{Erlich:2005qh}. The experimental data listed in the
last column  are taken from the particle data
group~\cite{Amsler:2008zz}. The empirical decay constants and coupling
constants are extracted from the corresponding decay
widths~\cite{Erlich:2005qh}.  All results are given  in units of MeV
except for the condensate and the ratio of two  condensates.}
 \label{tab:2}
\end{table}
%%%
Our results are summarized in table~\ref{tab:2}. We use the
$\rho$-meson mass as an input and do the global fit to the other
observables as in Ref.~\cite{Erlich:2005qh}.  We obtain three
different sets of the results for which we call Model I, II, and III.
For the sake of comparison, we also list the results of
ref.~\cite{Erlich:2005qh} as Model IV.  Note that the value of the
ratio between the quark condensate and mixed quark-gluon condensate,
{\em i.e.} $m_0^2$, has not been uniquely determined
\cite{Ioffe:1981kw,Belyaev:1982sa,Dosch:1988vv,Dorokhov:1997iv,
  Polyakov:1996kh,Doi:2002wk,Nam:2006ng}.  For example, while Belyaev
and Ioffe have predicted $m_0^2\simeq 0.8\,{\rm GeV^2}$, based on the
QCD sum rules~\cite{Belyaev:1982sa}, Doi et al. have obtained $m_0^2
\simeq 2.5\,\mathrm{GeV}^2$~\cite{Doi:2002wk} in quenched lattice QCD.

In the present study, it turns out that the value of $m_0^2$ is mostly
fixed by the mass of $a_1$.  The pion decay constant from Model
I seems to be quite underestimated in comparison with the data.
Moreover, the corresponding result of $m_0^2$ becomes much
larger than those of other works.  Thus, it implies that Model I seems
to be ruled out.  On the other hand, the results from
Model II and Model III are in qualitative agreement with measured
observables.  However, the values of $m_0^2$ are rather different each
other.

Since the ratio $m_0^2$ is sensitive to $a_1$ mass, we calculate a
coupling involving $a_1$ to see if our model can
select a particular value of $\sigma_M/\sigma$.  The coupling of
four $a_1$ fields can be determined in the following way:
\begin{eqnarray}
  \label{eq:1}
  S_{A4}^{4D} &=& g_{A4}{\rm Tr}\int d^4x A(x)_\mu A(x)_\nu A(x)_\mu
  A(x)_\nu\,,\cr
  g_{A4} &\equiv& 2\int_{\epsilon}^{z_m} dz\frac{1}{z}\psi_{a_1}(z)^4\, .
\end{eqnarray}
Since we have already calculated the $a_1$ meson wave function, it is
straightforward to read out the results for $g_{A4}$. The values of
$g_{A4}$ are listed in the last row of table~\ref{tab:2}.  As shown in
the table, the value of $g_{A4}$ is almost stable with $m_0^2$ varied.
However, since the $m_0^2$ from Model II is comparable to that of the
QCD sum rules~\cite{Belyaev:1982sa}, Model II is favored in the
present work.

\section{Summary and Conclusion}
In the present work, we have aimed at studying the quark-gluon mixed
condensate within the framework of the hard-wall AdS/QCD
model~\cite{Erlich:2005qh,DaRold:2005zs}. To this end, we have
introduced a bulk scalar field $\Phi$ dual to the mixed condensate to
the hard wall model.  We have treated the ratio of the chiral and
mixed condensates $m_0^2$ as a free parameter and fixed it by
phenomenology. Our results are summarized in table 2.  Model II in the
table predicts $m_0^2\simeq 0.72~{\rm GeV^2}$, which is comparable to
that from the QCD sum rules $m_0^2\simeq 0.8~{\rm
  GeV^2}$~\cite{Belyaev:1982sa}.  It indicates that Model II is the
most favorable one, though we were not able to uniquely
fix the value of the mixed condensate in the present work.  However,
the present simple hard-wall model seems to be not adequate to fix the
mixed condensate uniquely.   Nevertheless, we conclude that the mixed
condensate should be considered as an important ingredient of
low-energy QCD as well as the chiral condensate in any AdS/QCD models.

Finally, we remark that it will be interesting to see if one can study
the mixed condensate in a stringy set-up. This is due to the fact
that, unlike the chiral or gluon condensate, the mixed quark-gluon
condensate is associated with two completely different branes such as
$D3-D7$ or $D4-D8$.

%%%%%%%%%%
\acknowledgments
The present work is supported by Inha University Research Grant
(INHA-37453).

\end{document}